\documentclass[
 aip,
 amsmath,amssymb,
 reprint,%
]{revtex4-2}
\usepackage{xcolor}
\usepackage{graphicx}
\usepackage{dcolumn}
\usepackage{bm}

\usepackage[utf8]{inputenc}
\usepackage[T1]{fontenc}

\usepackage{etoolbox}
\usepackage{amsmath}
\usepackage{comment}
\usepackage{hyperref}
\usepackage[capitalise]{cleveref}
\usepackage{xcolor}
\hypersetup{
    colorlinks,
    linkcolor={red!50!black},
    citecolor={blue!50!black},
    urlcolor={blue!80!black}
}
\usepackage{ulem}


\newcommand{\PtxAu}{\ensuremath{ \mathrm{Pt}_{x}\mathrm{Au}_{100-x} }}

\begin{document}
\preprint{AIP/123-QED}

\title{High-efficiency Pt$_\mathrm{75}$Au$_\mathrm{25}$-based spintronic terahertz emitters}

\author{Wenlu Shi$^{\dagger}$}
\affiliation{\mbox{Department of Physics and Astronomy, University of California Irvine, Irvine, California 92697, USA}}

\author{Gene D. Nelson$^{\dagger}$}
\affiliation{\mbox{Department of Physics and Astronomy, University of California Irvine, Irvine, California 92697, USA}}

\author{Han-Hsuan Wu}
\affiliation{\mbox{Department of Physics and Astronomy, University of California Irvine, Irvine, California 92697, USA}}

\author{Yiwei Ju}
\affiliation{\mbox{Department of Materials Science and Engineering, University of California Irvine, Irvine, California 92697, USA}}

\author{Xiaoqing Pan}
\affiliation{\mbox{Department of Physics and Astronomy, University of California Irvine, Irvine, California 92697, USA}}
\affiliation{\mbox{Department of Materials Science and Engineering, University of California Irvine, Irvine, California 92697, USA}}
\affiliation{\mbox{Irvine Materials Research Institute, University of California Irvine, Irvine, California 92697, ~USA}}

\author{Wilson Ho}
\affiliation{\mbox{Department of Physics and Astronomy, University of California Irvine, Irvine, California 92697, USA}}
\affiliation{\mbox{Department of Chemistry, University of California Irvine, Irvine, California 92697, USA}}

\author{Ilya N. Krivorotov}

\email{ilya.krivorotov@uci.edu\\$\dagger$ These authors contributed equally.}
\affiliation{\mbox{Department of Physics and Astronomy, University of California Irvine, Irvine, California 92697, USA}}

\begin{abstract}
Spintronic terahertz emitters (STEs) generate broadband THz radiation via ultrafast spin–charge conversion in magnetic multilayers, offering spectral coverage beyond that of photoconductive antennas and nonlinear optical crystals. 
Here, we demonstrate STEs based on \PtxAu\, alloy that achieves significantly higher THz output power than widely used Pt-based devices.
Alloy composition and layer thickness tuning yield $\mathrm{Pt_{75}Au_{25}}$ as the optimal alloy providing a 30\,\% increase in THz power in CoFeB/$\mathrm{Pt_{75}Au_{25}}$ bilayer STEs compared to the optimized CoFeB/Pt reference STE. 
In W/CoFeB/$\mathrm{Pt_{75}Au_{25}}$ trilayer STEs, we observe a 10\,\% higher THz power than in the optimized W/CoFeB/Pt trilayer. 
The STE efficiency is reduced upon annealing for both $\mathrm{Pt_{75}Au_{25}}$- and Pt-based STEs due to formation of interfacial alloys. 
Our results establish $\mathrm{Pt_{75}Au_{25}}$ as a promising platform for high-performance STEs, where its giant spin Hall effect significantly enhances efficiency over conventional Pt-based devices.
\end{abstract}

\maketitle



Terahertz (THz) radiation has multiple applications in spectroscopy \cite{mootz_visualization_2022,wang_atomic-scale_2022,fava_magnetic_2024,park_capturing_2024}, imaging \cite{yamaguchi_brain_2016,valusis_roadmap_2021,yan_terahertz_2025,niwa_switchable_2021}, and ultrafast spintronics \cite{li_observation_2018,cheng_studying_2021,blank_empowering_2023,zhang_terahertz-field-driven_2024, kampfrath_resonant_2013,li_laser_2022,lezier_fully_2022, khymyn_antiferromagnetic_2017}. 
Despite its significance, efficient THz sources and detectors remain a key technological challenge.
A recent promising strategy for efficient THz generation is the use of spintronic THz emitters (STEs), which produce ultrashort THz pulses by converting laser-induced spin currents in magnetic multilayers into high-density charge currents and intense electromagnetic radiation via the inverse spin Hall effect (ISHE)\cite{seifert_efficient_2016,seifert_ultrabroadband_2017,seifert_spintronic_2022,wu_principles_2021,bull_spintronic_2021}. 
STEs based on nonmagnetic conductor/ferromagnet (NM/FM) multilayers with large ISHE in the NM, such as Pt/CoFeB bilayers and W/CoFeB/Pt trilayers, have shown efficient THz emission \cite{seifert_efficient_2016, seifert_spintronic_2022}. 
This efficiency arises from the giant spin Hall effect (SHE) in Pt \cite{kampfrath_terahertz_2013, huisman_femtosecond_2016,nguyen_spin_2016,duan_spin-wave_2014,evelt_spin_2018} and W \cite{hoffmann_spin_2013, seifert_efficient_2016}, which enables robust spin-to-charge conversion \cite{seifert_efficient_2016}.

Further development of more efficient spintronic THz emitters is essential to fully exploit their potential for compact, broadband, and high-power THz technologies.
Strategies for increasing STE efficiency include multilayer design \cite{feng_highly_2018}, thermal control \cite{vogel_average_2022,vaitsi_rotating_2024,gaoEnhancedSpintronicTerahertz2019b}, alloy engineering \cite{sasaki_effect_2019}, as well as interface engineering and optimization \cite{feng_highly_2018,sasaki_annealing_2017,torosyan_optimized_2018,rouzegar_broadband_2023,ren_effect_2024}. 
However, quantifying application-relevant improvements remains challenging: STE performance is highly sensitive to layer thicknesses, and each multilayer system must be optimized, particularly in trilayer geometries\cite{chengEfficientTemperatureindependentTerahertz2022,janusEmergentSpinHall2025, janusEnhancedTHzEmission2025,sasaki_effect_2019,scheuerTHzEmissionFe2022}. 

In this work, we explore the recently demonstrated giant SHE in \PtxAu\, alloy \cite{zhu_highly_2018} to develop efficient STEs that outperform optimized Pt-based devices.
We systematically investigate THz emission in both CoFeB/\PtxAu\ bilayers and W/CoFeB/\PtxAu\, trilayers and compare them to optimized Pt-based counterparts. 
By tuning the \PtxAu\, alloy composition and individual layer thicknesses, we find that CoFeB(1.6\,nm)/$\mathrm{Pt_{75}Au_{25}}$(3.0\,nm) bilayer achieves a 30\,\% enhancement in emitted THz power compared to the optimized CoFeB(1.6\,nm)/Pt(2.1\,nm). 
The optimized W(1.8\,nm)/CoFeB(1.3\,nm)/$\mathrm{Pt_{75}Au_{25}}$(3.0\,nm) trilayer exhibits a 10\,\% THz emission enhancement compared to the optimized W(1.8\,nm)/CoFeB(1.3\,nm)/Pt(2.1\,nm) STE.

\begin{figure*}[!t]
    \centering
    \includegraphics[width=1\textwidth]{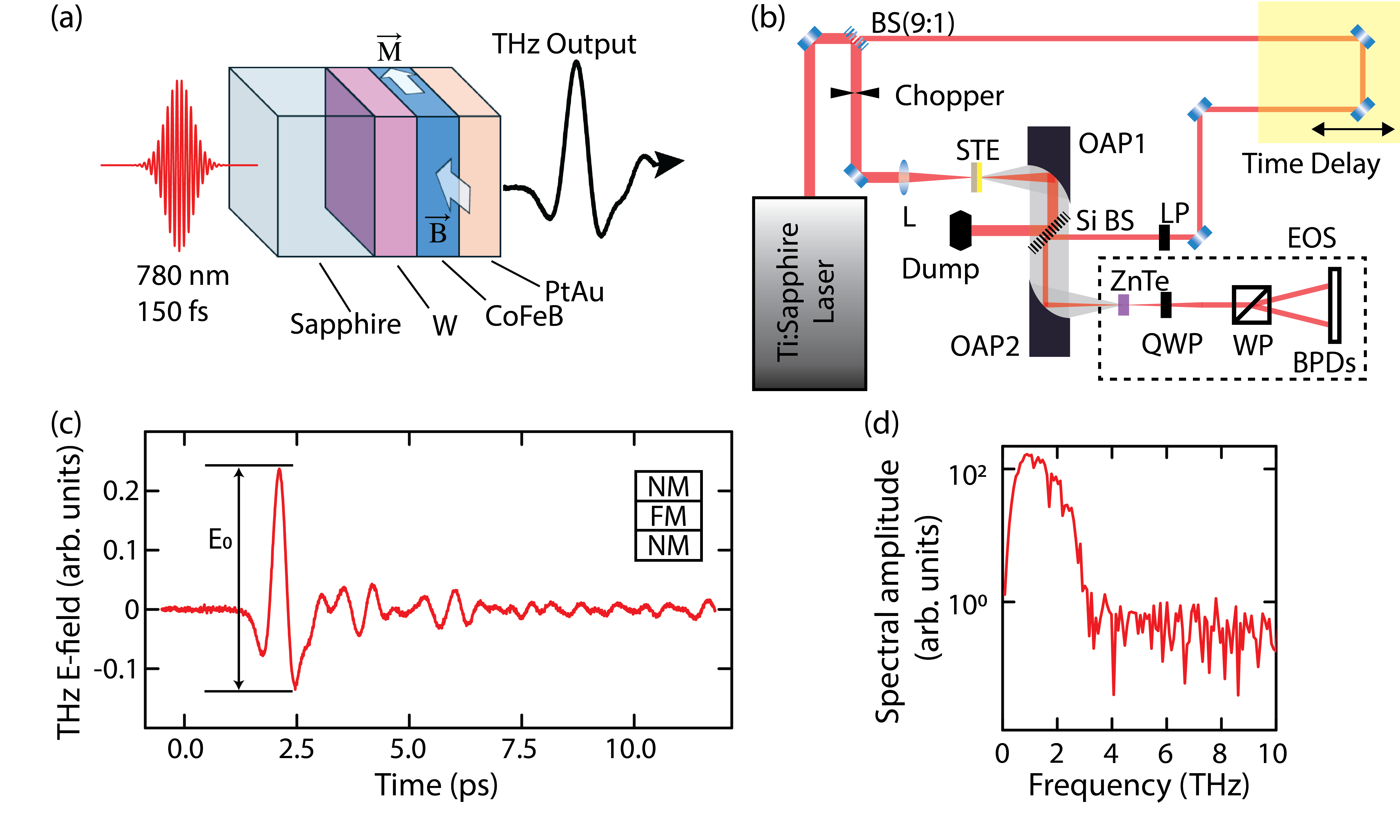}
    \caption{(a) Schematic of a W/CoFeB/\PtxAu\, STE multilayer on a sapphire substrate, where $\overrightarrow{B}$ is the external magnetic flux density, and $\overrightarrow{M}$ is the equilibrium magnetization of the FM layer. 
    (b) Experimental setup of THz electro-optic sampling (EOS) for STE THz emission measurements using a Ti:sapphire laser, including a beam splitter (BS), a chopper, a lens (L), a silicon beam splitter (Si BS), a beam dump (Dump), two off-axis parabolic mirrors (OAP1, OAP2), a linear polarizer (LP), a 500\,$\mu$m ZnTe crystal detector,  a quarter-waveplate (QWP), a Wollaston prism (WP), and a balanced photodetector (BPDS). 
    (c) Time-domain waveform of the emitted THz $E$-field  from the W(1.8\,nm)/CoFeB(1.3\,nm)/$\mathrm{Pt_{75}Au_{25}}$(3.0\,nm) trilayer STE measured via EOS. ${E_0}$ marks the peak-to-peak amplitude of the THz waveform.
    (d) Fourier transform of the THz waveform in (c) plotted as spectral amplitude versus frequency.}
    \label{fig:figure1}
\end{figure*}

The STE multilayers used in this study were deposited by magnetron sputtering at Ar process gas pressure of 2\,mTorr, with a base pressure maintained below $2\times{10}^{-8}$\,Torr. 
The samples were grown on 0.5\,mm thick single-crystal $\mathrm{Al_2O_3(0001)}$ (sapphire) substrates, with the multilayer structures of $\mathrm{Al_2O_3}$/CoFeB/\PtxAu\ for bilayers and $\mathrm{Al_2O_3}$/W/CoFeB/\PtxAu\ for trilayers as shown in \cref{fig:figure1} (a). 
We used an Fe-rich composition of CoFeB ($\mathrm{Co_{20} Fe_{60} B_{20}}$) to enhance the FM saturation magnetization \cite{fu_temperature_2016} and STE efficiency \cite{seifert_efficient_2016,sasaki_effect_2019}. The \PtxAu\, layer was grown by co-sputtering from Au and Pt targets.

Terahertz emission was characterized using a standard THz electro-optic sampling (EOS) setup as illustrated in \cref{fig:figure1}(b). 
A Ti:sapphire laser operating at 780\,nm with 150\,fs pulse duration and 80\,MHz repetition rate was used as the STE driving light source. 
The output beam was split into pump and probe paths. The pump beam was modulated using an optical chopper for phase-sensitive detection and focused onto the STE sample. 
The magnetization of the STE was saturated in the sample plane by an external magnetic field of 80\,mT. 
THz pulses generated from the sample were collected and directed onto a 500\,$\mu$m thick ZnTe EOS crystal detector. 
A high-resistivity float-zone silicon (HRFZ-Si) beam splitter (TYDEX) was employed to filter out the residual pump beam and direct the THz pulses toward the ZnTe crystal. 
The co-propagating probe beam was also redirected onto the ZnTe by the beam splitter. 
The THz-induced birefringence in the ZnTe crystal modulated the polarization state of the probe pulse, which was analyzed using a quarter-wave plate (QWP), Wollaston prism (WP), and balanced photodetector system (BPDS). 
Temporal waveforms of the THz electric field were acquired by scanning the optical delay stage, enabling ultrafast dynamic characterization of the emitted THz pulses.
\begin{figure*}[!t]
    \centering
    \includegraphics[width=1\textwidth]{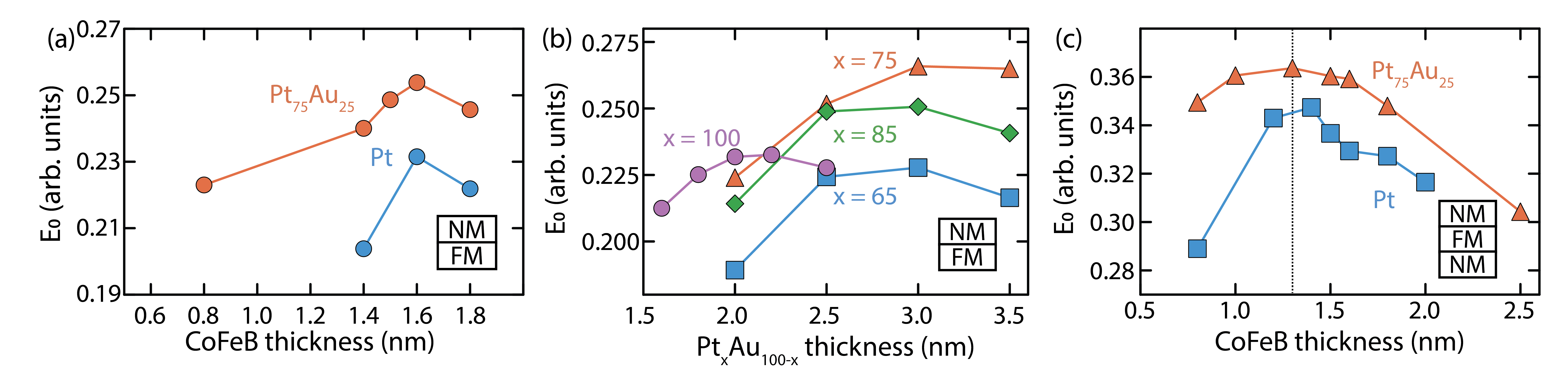}
    \caption{Optimization of THz emission with respect to \PtxAu\, and CoFeB layer thicknesses: (a)
   THz electric field amplitude $E_0$ as a function of CoFeB thickness in CoFeB($d_\mathrm{FM}$)/Pt(2.0\,nm) and CoFeB($d_\mathrm{FM}$)/$\mathrm{Pt_{75}Au_{25}}$(3.0\,nm) bilayers. 
    (b) $E_0$ as a function of \PtxAu\, thickness for four Pt concentrations: $x$ = 65, 75, 85, and 100 in CoFeB(1.6\,nm)/ \PtxAu($d_\mathrm{NM}$)\ bilayers. 
    $E_0$ increases with \PtxAu\, thickness up to a composition-dependent maximum, with $\mathrm{Pt_{75}Au_{25}}$ showing the highest overall $E_0$ at $d_\mathrm{NM}=3.0$\,nm. 
    (c) $E_0$ as a function of CoFeB thickness $d_\mathrm{FM}$ in W(1.8\,nm)/CoFeB($d_\mathrm{FM}$)/Pt(2.1\,nm) and W(1.8\,nm)/CoFeB($d_\mathrm{FM}$)/$\mathrm{Pt_{75}Au_{25}}$(3.0\,nm) trilayer STEs. 
    Both systems maximize THz emission at $d_\mathrm{FM} \approx$\,1.3\,–\,1.4\,nm, with the $\mathrm{Pt_{75}Au_{25}}$-based STEs outperforming the Pt-based STEs at any $d_\mathrm{FM}$.}
    \label{fig:figure3}
\end{figure*}

A typical time-domain waveform of the THz electric field produced by a W(1.8\,nm)/CoFeB(1.3\,nm)/ $\mathrm{Pt_{75}Au_{25}}$(3.0\,nm) trilayer STE is shown in \cref{fig:figure1}(c). 
The signal exhibits a pronounced single-cycle pulse followed by weaker oscillations, characteristic of broadband THz emission from spintronic sources. 
The sharp leading edge and well-defined zero-crossing confirm the ultrafast nature of the spin current dynamics and spin-to-charge conversion in the multilayer structure. 
The corresponding spectrum obtained via the fast Fourier transform of the data in \cref{fig:figure1}(c) is shown in \cref{fig:figure1}(d). 
The spectrum reveals a bandwidth extending up to 3\,THz, limited primarily by the ZnTe detection crystal response\cite{casalbuoni_numerical_2008}.
Small dips observed in the spectrum are attributed to atmospheric water vapor absorption, a common feature in ambient THz measurements \cite{Xin_Terahertz_2006}. 

We first examined bilayers to isolate the impact of \PtxAu\, alloy composition on the STE performance. 
\cref{fig:figure3}(a) shows $E_0$ as a function of CoFeB thickness $d_\mathrm{FM}$ in CoFeB($d_\mathrm{FM}$)/Pt(2.0\,nm) and CoFeB($d_\mathrm{FM}$)/$\mathrm{Pt_{75}Au_{25}}$(3.0\,nm) bilayers, respectively. In both cases, $E_0$ exhibits a maximum at $d_\mathrm{FM}=1.6$\,nm.

\cref{fig:figure3}(b) shows $E_0$ as a function of \PtxAu\, thickness $d_\mathrm{NM}$ for $x=65,\,75,\,85\ \text{and}\ 100$ with $d_\mathrm{FM}=1.6$\,nm. 
For all compositions, $E_0$ exhibits a maximum as a function of $d_\mathrm{NM}$: $E_0(d_\mathrm{NM})$ peaks at $d_\mathrm{NM}=2.1$\,nm for pure Pt and at $d_\mathrm{NM}=3.0$\,nm $\mathrm{Pt_{75}Au_{25}}$.  The $\mathrm{Pt_{75}Au_{25}}$ alloy composition yields the strongest overall THz emission.
\cref{fig:figure3}(c) reveals that the optimized CoFeB(1.6\,nm)/$\mathrm{Pt_{75}Au_{25}}$(3.0\,nm) bilayer yields a 15\% increase in peak $E_0$ relative to the optimized Pt reference, corresponding to a 30\% enhancement in THz power. 
These results are consistent with previous work on SHE studies in \PtxAu\, alloy \cite{zhu_highly_2018} that demonstrated the maximum spin Hall angle at $x=75$. 

A microscopic theory of STE \cite{foggetti_quantitative_2025, kefayati_deciphering_2025} is based on super-diffusive transport and inelastic scattering of spin-polarized hot electrons excited by the laser pulse in the FM layer. The predictions of this theory are necessarily numerical and go beyond the scope of this work.
For qualitative understanding of the dependence of $E_0$ on the NM and FM layer thicknesses in \cref{fig:figure3}, we employ phenomenological models of STE \cite{torosyan_optimized_2018,yang_theoretical_2024}.
In the framework of these models, $E_0(d_\mathrm{NM})$ exhibits a peak near $d_\mathrm{NM} \approx 2 \lambda_\mathrm{NM}$, where $\lambda_\mathrm{NM}$ is the spin diffusion length in the nonmagnetic layer. $E_0$ initially increases with $d_\mathrm{NM}$ approximately as $\tanh{(d_\mathrm{NM}/ 2 \lambda_\mathrm{NM})}$, because in this regime increasing $d_\mathrm{NM}$ allows the spin current injected from the FM into the NM to be converted more efficiently into charge current via the ISHE. 
For thicker NM layers, $E_0$ decreases with $d_\mathrm{NM}$, as the additional NM thickness well above  $\lambda_\mathrm{NM}$ enhances electromagnetic wave absorption and partially shunts the ISHE-driven charge current. 
Quantitative numerical calculations of $E_0(d_\mathrm{NM})$ also show a characteristic peak \cite{torosyan_optimized_2018,yang_theoretical_2024, kefayati_deciphering_2025,foggetti_quantitative_2025}.

\begin{figure*}[!t]
    \centering
    \includegraphics[width=1\textwidth]{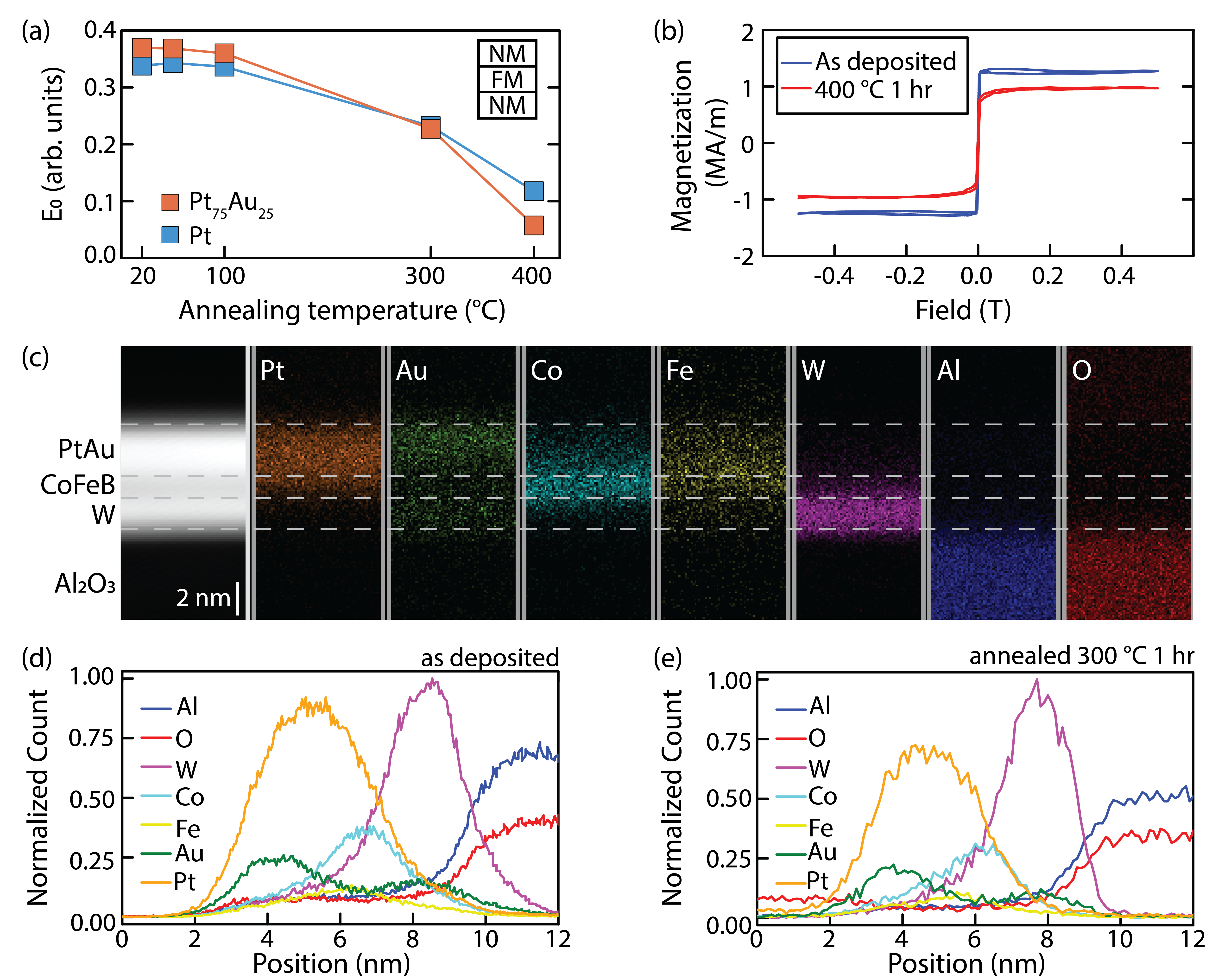}
    \caption{Effect of annealing on STE: (a) THz electric field amplitude $E_0$ as a function of annealing temperature for the optimized Pt- and PtAu-based trilayer STEs. 
    (b) Magnetization hysteresis loops of the W(1.8\,nm)/CoFeB(1.3\,nm)/ $\mathrm{Pt_{75}Au_{25}}$(3\,nm) STE before and after annealing at 400\,°C.
    (c) Cross-sectional STEM-EDS elemental maps of the W(1.8\,nm)/CoFeB(1.3\,nm)/$\mathrm{Pt_{75}Au_{25}}$(3\,nm) STE annealed at 300\,°C, showing spatial distributions of Pt, Au, Co, Fe, W, Al, and O. 
    (d, e) Normalized EDS line profiles along the film growth direction for (d) as-deposited and (e) 300\,°C-annealed samples. }
    \label{fig:figure4}
\end{figure*}
The larger optimal thickness of $\mathrm{Pt_{75}Au_{25}}$ in the STE compared to the Pt-based STE in \cref{fig:figure3}(c) is consistent with the spin diffusion lengths previously determined from spin Hall torque measurements: $\lambda_\mathrm{Pt_{75}Au_{25}} = 1.7$\,nm and $\lambda_\mathrm{Pt} = 1.4$\,nm \cite{zhu_highly_2018,nguyen_spin_2016}. 
Additionally, the higher observed THz field amplitude in the $\mathrm{Pt_{75}Au_{25}}$-based STE is consistent with the higher spin Hall efficiency of $\mathrm{Pt_{75}Au_{25}}$ relative to Pt \cite{zhu_highly_2018,nguyen_spin_2016}. 
The maximum $E_0$ observed in \cref{fig:figure3}(b) for the $\mathrm{Pt_{75}Au_{25}}$-based STE is also consistent with previous studies demonstrating that the maximum spin Hall efficiency in the \PtxAu alloy is found for $x=$\,75 \cite{zhu_highly_2018}.
The data in \cref{fig:figure3}(b) illustrate that both the alloy composition and the layer thickness must be carefully optimized in tandem to maximize the THz emission from spintronic heterostructures.

Having established $\mathrm{Pt_{75}Au_{25}}$ as the optimal \PtxAu \, alloy composition, we turn to trilayer STEs to further boost the emitted THz power. To this end, we introduced a W layer under the CoFeB/$\mathrm{Pt_{75}Au_{25}}$\ stack, as the large negative spin Hall angle in $\beta$-W is known to enhance STE efficiency \cite{haoBetaTungstenThin2015, seifert_ultrabroadband_2017, seifert_efficient_2016}.
For trilayer optimization, the Pt and $\mathrm{Pt_{75}Au_{25}}$ layer thicknesses were fixed at their optimal bilayer values.
First, we varied the W thickness in $\mathrm{Pt_{75}Au_{25}}$-based trilayers at fixed CoFeB thickness  ${d_\mathrm{FM} = 1.6}$\,nm and found the emitted THz power to be only weakly dependent on $d_\mathrm{W}$, with a broad maximum near $d_\mathrm{W}=1.8$\,nm consistent with prior studies \cite{seifert_efficient_2016,fengspintronicTerahertzEmitter2021,seifert_ultrabroadband_2017,herapathImpactPumpWavelength2019,nandiAntennacoupledSpintronicTerahertz2019}.
The addition of W underlayer enhances $E_0$ in Pt-based STEs by up to ~50\,\%, whereas $\mathrm{Pt_{75}Au_{25}}$-based trilayers show a smaller gain in $E_0$ of ~37\,\%. 

Second, we optimized the CoFeB thickness $d_{\mathrm{FM}}$ for the trilayer geometry. 
\cref{fig:figure3}(c) shows $E_0(d_{\mathrm{FM}})$ for W(1.8\,nm)/CoFeB($d_{\mathrm{FM}}$)/$\mathrm{Pt_{75}Au_{25}}$(3.0\,nm) and W(1.8\,nm)/CoFeB($d_{\mathrm{FM}}$)/Pt(2.1\,nm) trilayers. Both structures exhibit a clear maximum, occurring at $d_\mathrm{FM}=1.3$\,nm for the $\mathrm{Pt_{75}Au_{25}}$-based and $d_\mathrm{FM}=1.4$\,nm for Pt-based trilayers. 
The initial increase of $E_0$ with $d_{\mathrm{FM}}$ reflects enhanced spin current generation as more spin-polarized carriers are produced in a thicker ferromagnetic layer. 
Increasing $d_\mathrm{FM}$ beyond the optimum reduces $E_0$, likely due to spin relaxation within the FM layer, which limits the spin current reaching the FM/NM interface \cite{yang_modeling_2023}. 
Across the $d_\mathrm{FM}$ range studied, $\mathrm{Pt_{75}Au_{25}}$-based STEs consistently exhibit stronger THz emission than Pt-based STEs. In summary, the optimized PtAu-based stack W(1.8\,nm)/CoFeB(1.3\,nm)/$\mathrm{Pt_{75}Au_{25}}$(3.0\,nm) yields 
a 10\,\% enhancement of the generated THz power compared to the optimized Pt-based W(1.8\,nm)/CoFeB(1.4\,nm)/ Pt(2.1\,nm) STE. These results highlight the importance of co-optimizing the FM and NM layers to maximize spin-to-charge conversion efficiency and THz output of STEs.

\begin{figure*}[!t]
  \centering
  \includegraphics[width=1\textwidth]{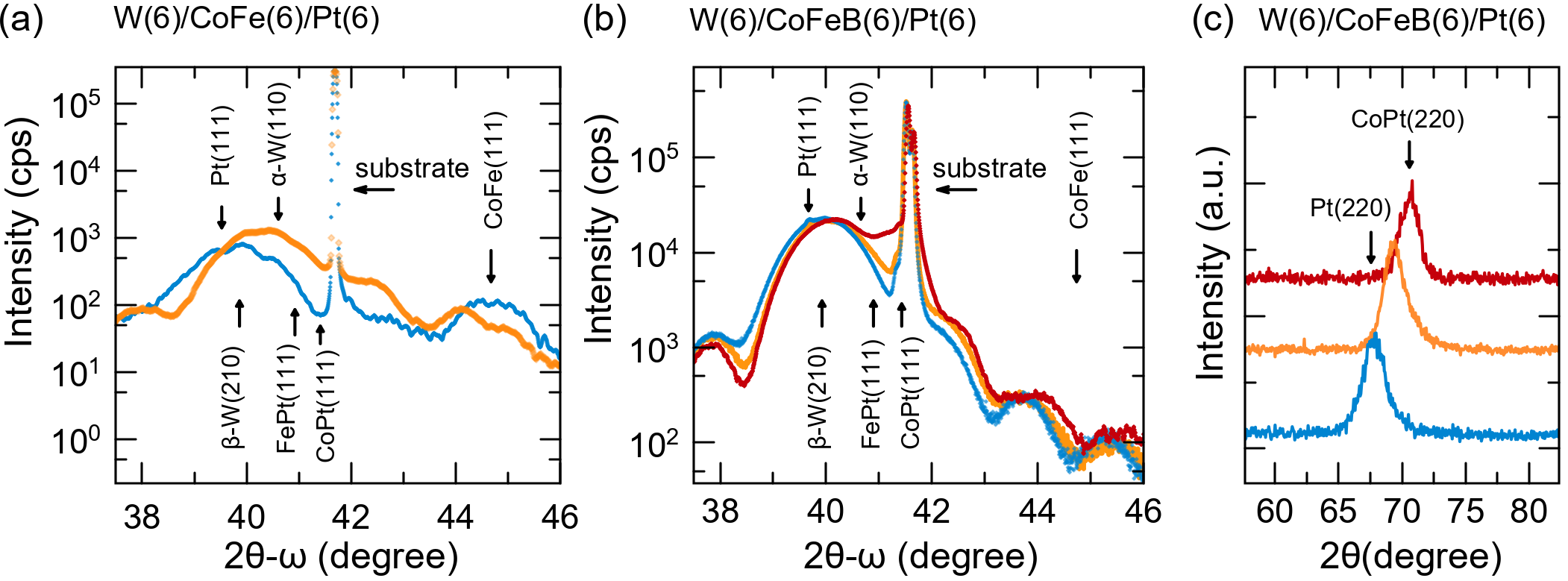}

  \caption{
  X-ray diffraction (XRD) $2\theta$--$\omega$ scans of 
  W(6\,nm)/Co$_{30}$Fe$_{70}$(6\,nm)/Pt(6\,nm) and 
  W(6\,nm)/CoFeB(6\,nm)/Pt(6\,nm) multilayers.
  (a) CoFe-based sample in the as-deposited state (blue) and after annealing at 
  $300^{\circ}\mathrm{C}$ for 3 hours (orange). 
  (b) CoFeB-based sample in the as-deposited state (blue), after annealing at 
  $300^{\circ}\mathrm{C}$ for 3 hours (orange), and after annealing at $400^{\circ}\mathrm{C}$ for 2 hours (red). 
  Arrows mark the expected Bragg peak positions of Pt, $\alpha$-W, $\beta$-W, FePt, CoPt, Fe$_3$Pt and CoFe. 
  (c) Grazing-incidence XRD ($\omega = 0.5^{\circ}$) of the CoFeB-based sample showing the 
  Pt(220) and CoPt(220) peaks. 
  Blue, orange, and red curves correspond to the as-deposited, $300^{\circ}\mathrm{C}$ annealed for 3 hours, and $400^{\circ}\mathrm{C}$ annealed for 2 hours samples, respectively.
  }
  
  \label{fig:singlecol}
\end{figure*}

Finally, to assess the thermal stability of the STEs, we investigate the influence of annealing on STE emission efficiency. 
For these studies, we anneal the optimized Pt- and $\mathrm{Pt_{75}Au_{25}}$-based STE trilayers W(1.8\,nm)/CoFeB(1.3\,nm)/$\mathrm{Pt_{75}Au_{25}}$(3.0\,nm) and W(1.8\,nm)/CoFeB(1.4\,nm)/Pt(2.1\,nm) under $1\times 10^{-6}$\,Torr vacuum for 1 hour. 
The samples were annealed at 4 different temperatures $T_a$ ranging from 50\,$^\circ$C to 400\,$^\circ$C. The Curie temperature of CoFeB films lies in the range of 480–630\,$^\circ$C, well above the maximum annealing temperature used in this study \cite{lee_temperature_2017}.

\Cref{fig:figure4}(a) shows that the THz emission amplitude $E_0$ decreases with increasing annealing temperature. To elucidate the effect of annealing on the magnetic properties of the multilayers, we made vibrating sample magnetometry measurements of W(1.8\,nm)/ CoFeB(1.3\,nm)/$\mathrm{Pt_{75}Au_{25}}$(3\,nm) STEs and found that saturation magnetization decreased by 15\,\% upon annealing at 400\,$^\circ$C, as shown in \cref{fig:figure4}(b). 
This demonstrates that annealing reduces the CoFeB magnetization, likely due to interlayer diffusion at the NM/FM interfaces and/or the formation of nonmagnetic or weakly magnetic interfacial alloys. 

To better understand the mechanism of the observed THz emission suppression upon annealing, we made scanning transmission electron microscopy – energy dispersive X-ray spectroscopy (STEM-EDS) measurements shown in \cref{fig:figure4}(c–e). 
Cross-sectional STEM specimens were prepared using a standard focused ion beam (FIB) lift-out procedure, and STEM-EDS elemental maps were acquired with a JEOL Grand ARM300F microscope operated at 300 kV.
The cross-sectional STEM-EDS images in \cref{fig:figure4}(c) show significant overlap of Pt, Au, Co, Fe and W spatial profiles, which is indicative of short-scale interfacial roughness and/or interdiffusion of these elements \cite{li_thz_2019}.
\Cref{fig:figure4}(c-e) reveals a noticeable segregation of Au in the W layer in both as-deposited and annealed samples.

Our STEM-EDS measurements reveal that the spatial atomic distribution of the elements is not strongly modified by annealing. This leaves us with two possible origins of the observed annealing-induced reduction of STE efficiency. 
First, annealing is known to induce diffusion of boron from CoFeB into adjacent layers \cite{drobitchEffectCoFeDusting2019,ghoshProbingSpinHall2022,greerObservationBoronDiffusion2012,saravananObservationUniaxialMagnetic2022,mahendraRoleInterfaceIntermixing2023,barsukov_field-dependent_2014}, which  cannot be seen in \cref{fig:figure4}(c) due to the poor STEM-EDS sensitivity to B, as its low K-shell X-ray energy makes it undetectable using this technique.
It is thus possible that B diffusion into W and $\mathrm{Pt_{75}Au_{25}}$ can decrease their spin Hall efficiencies and the FM/NM interfacial transparency to spin current, which can reduce STE efficiency \cite{drobitchEffectCoFeDusting2019}. 
Second, annealing can form alloys at the FM/NM interfaces, a process masked in STEM-EDS images due to significant interfacial roughness of the multilayers. 
Formation of nonmagnetic or weakly magnetic interfacial alloys can also decrease spin Hall efficiency and interfacial spin transparency \cite{ghoshProbingSpinHall2022}, resulting in reduced STE efficiency.

In order to study formation of interfacial alloys, we made X-ray diffraction (XRD) measurements of a W(6\,nm)/Co$_{30}$Fe$_{70}$(6\,nm)/Pt(6\,nm) and W(6\,nm)/CoFeB(6\,nm)/Pt(6\,nm) trilayer as-deposited and annealed at $300^{\circ}\mathrm{C}$ and $400^{\circ}\mathrm{C}$, shown in \cref{fig:singlecol}. 
The thicknesses of the layers were increased in comparison to those of the STE samples to be reliably measurable by the Rigaku SmartLab thin film diffractometer. We first study the CoFe-based samples, as CoFe alloys typically grow as rough polycrystalline films, suppressing Laue oscillations that would otherwise obscure weak diffraction features. The $2\theta$--$\omega$ scan in \cref{fig:singlecol}(a) reveals the presence of pure Pt, $\alpha$-W, $\beta$-W and a CoFe alloy in the as-deposited CoFe-based sample, suggesting minimal interfacial alloying.
To monitor annealing-induced phase transitions, we mark the angular positions of Bragg peaks of the as-deposited phases and possible interfacial alloys in \cref{fig:singlecol}. The data reveal that upon annealing, the contributions to the XRD intensity from Pt, $\beta$-W  and CoFe decrease while the contributions from FePt, CoPt, Fe$_3$Pt and $\alpha$-W increase. This is indicative of Pt-based interfacial magnetic alloy formation \cite{ristauRelationshipHighCoercivity1999,kimInfluenceCappingLayers2023,TsyntsaruIron-tungsten2009,kimEnhancementInterfacialDzyaloshinskii2023} and the $\beta$-W to $\alpha$-W phase transition induced by annealing. 
The interfacial magnetic alloys can reduce spin transparency of the NM/FM interface and thereby decrease STE output power.

We next repeat the experiment on CoFeB-based samples, as shown in \cref{fig:singlecol}(b). 
The $2\theta$--$\omega$ scans of these samples display clear Laue oscillations, consistent with the atomically smooth, amorphous growth of CoFeB. 
The annealing-induced changes in the XRD intensity are also consistent with formation of the Pt-based interfacial magnetic alloys\cite{ristauRelationshipHighCoercivity1999}. 
In the grazing-incidence XRD scan geometry (Fig.~\ref{fig:singlecol}(c)), the Pt(220) Bragg peak in the as-deposited sample shifts to higher $2\theta$ angles upon annealing, eventually coinciding with the CoPt(220) Bragg peak position, which provides an independent confirmation of Pt-based interfacial magnetic alloys.
These results agree with similar studies of alloying in annealed CoFeB-based multilayers \cite{kimInfluenceCappingLayers2023,latteryLowGilbertDamping2018,zhuThermalStabilityCoFeB2012,mahendraRoleInterfaceIntermixing2023,zhaoEnhancingDomainWall2019}. 

Therefore, XRD data in \cref{fig:singlecol} suggest that the reduced STE efficiency after annealing in our samples is the result of interfacial alloying in the STE multilayer. 
The fact that magnetization of the CoFeB layer in \cref{fig:figure4}(c) is reduced upon annealing also suggests loss of a fraction of Co and Fe atoms to CoPt and FePt interfacial alloys with reduced magnetization\cite{tsyntsaruStructuralMagneticMechanical2013,mahendraRoleInterfaceIntermixing2023}. 
A simple annealing-induced diffusion of boron out of the CoFeB layer typically increases its magnetization \cite{belmeguenai_Investigation_2018,drobitchEffectCoFeDusting2019,saravananObservationUniaxialMagnetic2022}. 

The data in \cref{fig:singlecol} indicate that upon annealing $\beta$-W partially transforms into the thermodynamically stable $\alpha$-W phase characterized by substantially reduced resistivity and diminished spin Hall angle relative to $\beta$-W\cite{haoBetaTungstenThin2015,petroffMicrostructureGrowthResistivity1973,nguyen_spin_2016}. 
Along with interfacial magnetic alloy formation, this can lead to reduced STE efficiency upon annealing.
Previous studies reported that annealing a glass/W/CoFeB/Pt trilayer STE did not enhance the THz output while glass/CoFeB/Pt and glass/CoFeB/W bilayer STEs exhibited increased output after annealing \cite{gaoEnhancedSpintronicTerahertz2019b}.
This behavior is consistent with the rougher CoFeB morphology on a W underlayer being more susceptible to interfacial alloying upon annealing \cite{zhu2018, aurongzeb2005}, compared to the smoother CoFeB films deposited directly on glass \cite{saravanan2025}.
We attribute the reduced STE output after annealing in our $\mathrm{Al_2O_3}$/W/CoFeB/$\mathrm{Pt_{75}Au_{25}}$ and $\mathrm{Al_2O_3}$/W/CoFeB/Pt trilayers to enhanced interfacial alloying, which is promoted by the increased roughness of W and CoFeB layers grown on crystalline $\mathrm{Al_2O_3}$ substrates compared to amorphous glass \cite{gaoEnhancedSpintronicTerahertz2019b}.
Prior studies \cite{sasaki_effect_2019} showed that using a smoother Ta layer instead of W yields enhanced THz emission after annealing in Ta/CoFeB/Pt trilayer STEs.

In conclusion, we have shown that alloy engineering provides a powerful route to enhance the efficiency of spintronic THz emitters. Among the \PtxAu\, alloys, $\mathrm{Pt_{75}Au_{25}}$ emerges as the optimal composition for STEs due to its highest spin Hall efficiency. 
Systematic layer thickness optimization reveals that the CoFeB(1.6\,nm)/$\mathrm{Pt_{75}Au_{25}}$(3.0\,nm) bilayer delivers 30\,\% more THz power than the optimized CoFeB(1.6\,nm)/Pt(2.1\,nm) Pt-based benchmark, while W(1.8\,nm)/CoFeB(1.3\,nm)/$\mathrm{Pt_{75}Au_{25}}$(3.0\,nm) trilayers yield 10\,\% higher THz power than the optimized W(1.8\,nm)/CoFeB(1.4\,nm)/Pt(2.1\,nm) Pt-based trilayer STE. 
These results establish $\mathrm{Pt_{75}Au_{25}}$ as a promising platform for high-performance spintronic THz emitters.
Further improvements of $\mathrm{Pt_{75}Au_{25}}$-based STEs may employ strategies recently demonstrated to significantly enhance THz signal from  Pt-based STEs, namely partial oxidation of the NM layer \cite{li_significant_2023} and doping the NM layer with MgO \cite{wang_enhancement_2023}.

This work was primarily supported by the National Science Foundation Materials Research Science and Engineering Center program through the UC Irvine Center for Complex and Active Materials (DMR-2011967). 
The development of the THz detection setup was supported by the Office of Basic Energy Sciences of the U.S. Department of Energy Award Number DE-SC0024037.
Partial support from the National Science Foundation via awards ECCS-2213690 and DMREF-2324203 as well as from the University of California National Laboratory Fees Research Program is also acknowledged.
We also thank the Eddleman Quantum Institute (EQI) for providing materials and supplies.
The authors acknowledge the use of facilities and instrumentation at the UC Irvine Materials Research Institute (IMRI), which is supported in part by the National Science Foundation through the UC Irvine Materials Research Science and Engineering Center (DMR-2011967).\\

\noindent \textbf{AUTHOR DECLARATIONS}

\noindent \textbf{Conflict of Interest}

\noindent The authors do not have conflicts of interest to disclose.\\

\noindent \textbf{Author Contributions}

\noindent W.S. developed THz measurement setup and made time-domain measurements of STE signals. G.D.N. deposited STE multilayers, made magnetometry and X-ray diffraction measurements. H.-H.W. and Y.J. made STEM and EDS measurements. I.N.K, W.H. and X.P. conceived and supervised the project. G.D.N., W.S. and I.N.K. analyzed the data and wrote the manuscript. All authors reviewed and edited the manuscript.\\

\noindent \textbf{DATA AVAILABILITY}

\noindent The data that support the findings of this study are available from the corresponding author upon reasonable request.

\bibliographystyle{aipnum4-2}  
\bibliography{main}
\end{document}